# Transformer Based Tissue Classification in Robotic Needle Biopsy

Fanxin Wang, Yikun Cheng, Sudipta S Mukherjee, Rohit Bhargava, and Thenkurussi Kesavadas

*Abstract*—Image-guided minimally invasive robotic surgery is commonly employed for tasks such as needle biopsies or localized therapies. However, the nonlinear deformation of various tissue types presents difficulties for surgeons in achieving precise needle tip placement, particularly when relying on low-fidelity biopsy imaging systems. In this paper, we introduce a method to classify needle biopsy interventions and identify tissue types based on a comprehensive needle-tissue contact model that incorporates both position and force parameters. We trained a transformer model using a comprehensive dataset collected from a formerly developed robotics platform, which consists of synthetic and porcine tissue from various locations (liver, kidney, heart, belly, hock) marked with interaction phases (pre-puncture, puncture, post-puncture, neutral). This model achieves a significant classification accuracy of 0.93. Our demonstrated method can assist surgeons in identifying transitions to different tissues, aiding surgeons with tissue awareness.

## I. Introduction

The diagnostic procedure for numerous diseases in solid tissues, including most cancers and cardiomyopathies, involves tissue biopsies. However, the presence of sampling errors and reduced accuracy in needle insertion during biopsies can impose constraints on precision sampling. While generally accurate, biopsy procedures could miss representative parts of the disease by this imprecision and may also result in adverse events such as internal bleeding [1], patient discomfort, and the potential for the seeding cancer cells [2].

Researchers in advanced imaging and surgical robotic biopsy development have made significant impacts on methods of precise needle insertion with the support of imaging systems [3]. However, imaging systems often have limitations in detecting subtle tissue variations due to inherent limitations in resolution and sensitivity. An alternate method to add information is available by integrating knowledge about the mechanical properties of various anatomical sites. The measurement of mechanical properties can also aid in understanding disease-induced changes and improving biopsy precision, using changes that are specific to diseases [4]. For instance, in chronic liver diseases, fibrosis can lead to an increase in the stiffness of the liver compared to normal tissues [5]. Researchers have explored various approaches to image the mechanical properties of tissues. Most of these elasticity imaging methods involve applying some form of stress or mechanical excitation to the tissue, measuring how the tissue responds to this stimulus, and using this response to calculate parameters of mechanical characteristics. Both ultrasound elastography and magnetic resonance elastography have the capability to noninvasively evaluate tissue stiffness, and they have already demonstrated their clinical utility as diagnostic tools for assessing hepatic fibrosis [6].

While elastography has been employed to evaluate the overall stiffness of the liver, for real-time decision-making during procedures, surgeons still rely on their intuition to measure variances in tissue properties while performing manual needle interventions. They employ experience-built expertise to identify anatomical structures and events, including puncture occurrences and the differentiation between soft and hard tissues. Their perception is made possible through the haptic feedback derived from the interaction between the needle and the tissue. In breast cancer [7], for example, it was asserted that the mean shear stiffness of breast carcinoma is approximately 418% higher than the mean value of the surrounding tissue, whose resistance force change would be intuitively sensed by surgeon during the needle operation. Researchers have utilized the interaction force between the needle and the tissue to develop classifiers for distinguishing tissue properties. Statistical models and learning-based models are used for classification for liver needle insertion based on force patterns [8], Young's modulus is identified based on the energy stored in the needle-tissue system [9], a recurrent neural network (RNN) based Long-Short Term Memory (LSTM) model was trained to estimate the various synthetic tissue classes [10]. These studies validate the concept that the integration of a force parameter with position data can enhance the comprehension of

Fanxin Wang, is with Department of Mechatronics and Robotics, Xi'an Jiaotong-Liverpool University, 215123 China. (e-mail: Fanxin.Wang@xjtlu.edu.cn).
Yikun Cheng, is with Department of Mechanical Science and Engineering, University of Illinois Urbana-Champaign, Champaign, IL 61801 USA. (e-mail: yikun2@illinois.edu).
Sudipta S Mukherjee is with Beckman Institute for Advanced Science and Technology and Departments of Bioengineering, Electrical & Computer Engineering, Mechanical Science & Engineering, Chemical and Biomolecular Engineering, University of Illinois Urbana-Champaign, Champaign, IL 61801 USA. (e-mail: sudiptam@illinois.edu).
Rohit Bhargava is with Beckman Institute for Advanced Science and Technology, University of Illinois Urbana-Champaign, Champaign, IL 61801 USA. (e-mail: rxb@illinois.edu).
Thenkurussi Kesavadas, is with Division of Research and Economic Development, University at Albany - State University of New York, Albany, NY 12222 USA. (e-mail: tkesavadas@albany.edu).

interactions in a wide range of interventional procedures and diagnostic processes. However, the overall accuracy in tissue classification is still not satisfying, especially dealing with multiple kinds of tissues [13].

This study reports on creating a tissue classification algorithm aimed at enhancing the accuracy of robotic systems in needle biopsy, especially on detecting transitions between different types of tissues. In this paper, we report on tissue modeling, focusing on mechanical properties, dividing the insertion procedure into three distinct phases to identify the mechanics of needle-tissue reaction force, along with the experimental setup, data acquisition, enrichment, and labeling procedures. This modeling not only strengthens the following up classification in single type of tissue, but also enhances the stacking tissues classification in abdominal biopsy. We then implemented a transformer classification model, applying its superior ability to handle long-range dependencies in comparison with convolutional neural network (CNN) and recurrent neural network (RNN). Finally, we demonstrated online recognition of tissue reaction force as time series signals, proposing a tissue recognition framework that incorporates the classification model with real-time updates of normal tissue properties during insertion. This tissue detection offers a promising approach for robust and precise environment sensing, with broad applications in haptic biopsy sensing, ultimately enhancing the safety and effectiveness of medical procedures.

## II. Models and Methods

### A. Mechanics of Needle-tissue Interaction

To detect transitions in mechanical properties of diseased tissue, we formalized a mechanical model to describe the transitions in needle-tissue interactions based on needle insertion. These interactions have been modeled in the context of various medical procedures [14], which are summarized well in a survey [15]. Advancements encompass the modeling of cutting forces during insertion, the study of tissue material deformation, analysis of needle deflection throughout the needle insertion process, and the development of robot-controlled insertion procedures. An influential study [16] provided insight into the modeling of needle insertion forces when applied to bovine liver tissue. This study effectively characterized the insertion process into three distinct phases, with the puncture of the tissue capsule marking a key transition point.

The insertion force is a summation of stiffness, friction and cutting force:

$$f_{needle}(x) = f_{stiffness}(x) + f_{friction}(x) + f_{cutting}(x) \quad (1)$$

where $x$ stands for the insertion distance. The stiffness force occurs before puncture of the capsule, and the friction and cutting forces occur after this main puncture. Stiffness is best fit by a second-order polynomial of the form, friction force is modelled by modified Karnopp model [16], and cutting force is considered as constant for a given tissue. This research marks the key transition point at the puncture event and divides the insertion process into three distinct phases: pre-puncture, puncturing, and post-puncture.

For our model of needle-tissue interaction, we used the method of Okamura [16] for separating the insertion process into three distinct phases. Figure 1(a) illustrates the collected data samples within this insertion phase model. In pre-puncture phase, needle interaction force is dominated by stiffness and increases with depth. In the puncture phase, as the needle penetrates the surface, the reaction force curve experiences a sudden decrease, attributed to the decline of the stiffness force. Simultaneously, cutting force emerges, contributing to the overall needle-tissue reaction force. In post-puncture phase, friction force becomes evident, and when combined with the stationary cutting force, results in an overall increase in the needle interaction force as the needle travels deeper.

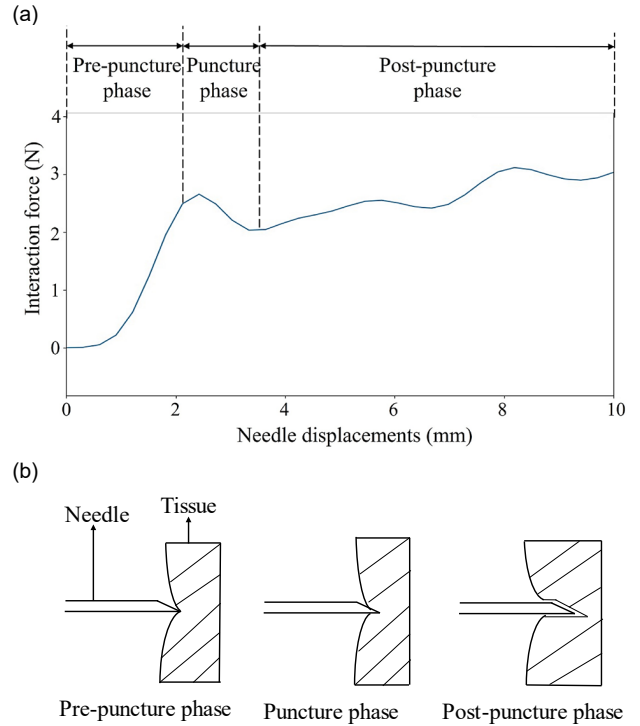

Fig. 1. Needle insertion phase model (a) Needle insertion phase on a sample data frame (b) Conceptual drawing of phase change

To notice, utilizing this mechanical modeling method, the process of consecutive tissue penetration in regions such as the breast or abdomen can be conceptualized as analogous to the stacking of needle insertions during a biopsy. This approach allows for the systematic separation of tissues, which is a crucial step leading to accurate tissue classification.

Based on these changes, figure 1(b) offers the corresponding conceptual illustration of the phase transitions occurring before, during, and after puncture takes place, where the insertion needle is represented by the thin object on the left, and the target tissue is represented by the solid block object on the right. Tissue deformation occurs primarily due to stiffness from pre-puncture phase. In the puncture phase, the needle pierces the surface, resulting in a partial cut into the tissue. In the post-puncture phase, the needle is fully

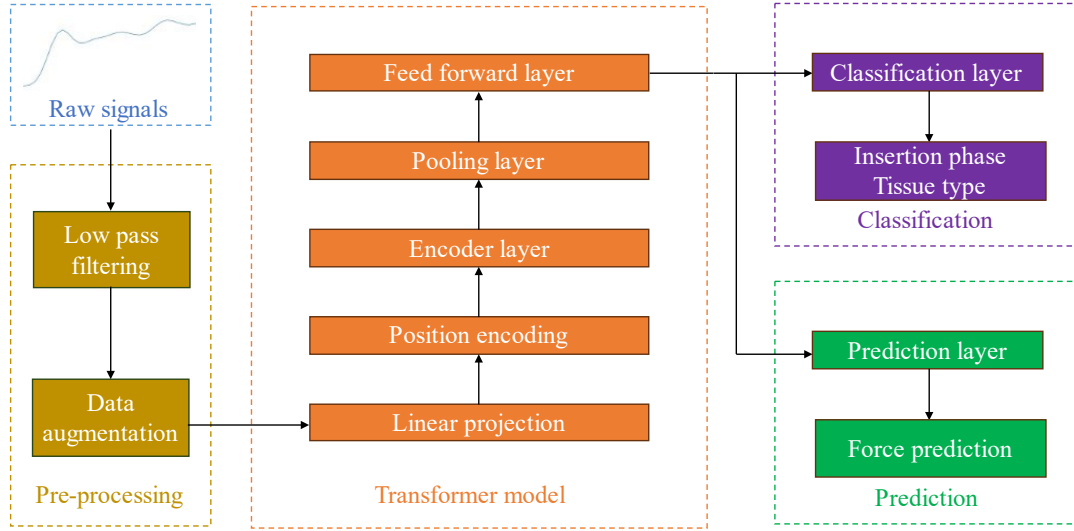

Fig. 2. Model architecture for classification

inserted into the tissue, with both friction force and cutting force being applied to the needle.

*B. Transformer Model for Classification*

While it is possible to develop analytical models, the complexities arising from variations in human anatomy, biomechanical properties, physiology, and geometry make the modeling process intricate and challenging to validate. To enhance model accuracy, a Finite Element Method (FEM) may be employed to simulate tissue deformation during needle insertion and analyze the interaction force [17]. FEM is precise when modeling small, linear elastic deformations. However, FEM calculations can be time-consuming, and the accuracy of FEM is highly reliant on the quality of its inputs.

In the context of these intricate modeling challenges, as an alternative to deterministic modeling, researchers have explored the application of stochastic machine learning techniques to assess the feasibility of comprehending needle-tissue interactions. A dichotomizer Bayesian classifier [9] was used to detect hepatic tissue and vein. A recurrent neural network (RNN) based Long-Short Term Memory (LSTM) model was trained to estimate the various synthetic tissue classes [10].

In recent years, the transformer model has garnered significant attention owing to its robust temporal modeling capabilities. It has found successful applications in diverse domains such as speech recognition [18] and computer vision [19]. The key innovation of the transformer model lies in its self-attention mechanism, which facilitates interactions among data points in the input sequence by computing similarity scores (attention weights) among them [20]. Furthermore, in comparison to recurrent neural networks (RNNs) like LSTM, which possess similar global context aggregation capabilities, transformers offer the advantage of parallel computation [21].

To detect tissue transitions, we developed a transformer-based model as illustrated in Figure 2. In recent years, transformer-based classification methodology not only serves greatly in engineering field [11], but also holds a great promise in recognizing medical field such as electroencephalogram (EEG) signals [12]. Our model architecture is depicted in Figure 2. It begins with the collection of position-force signals from various needle-tissue insertion experiments. The unprocessed signals include undesirable high-frequency noises that can adversely affect subsequent processing. Thus, a low-pass filter is needed to eliminate high-frequency noise. Data augmentation techniques are then applied to enrich the dataset, which will be discussed in the following section.

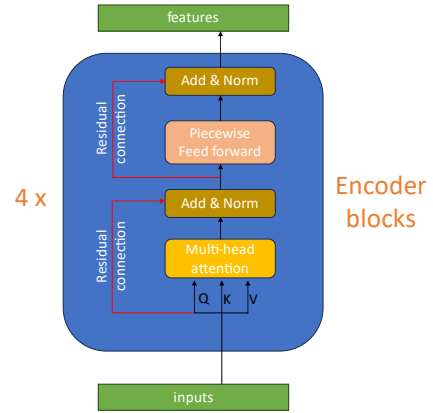

Fig. 3. Encoder layer details

The processed position-force signals pass through linear projection and position encoding steps to generate the necessary inputs for the transformer. An encoder layer is then employed to extract relevant features [28], as depicted in Figure 3. The encoder layers employ a self-attention mechanism within a blue box to enhance each token (embedding vector) by incorporating contextual information from the entire time series. Eight heads with size of 512 facilitate parallel attention calculations, granting access to diverse embedding subspaces. Within the encoder block, a position-wise feed-forward network (FFN) independently processes each embedding vector through a linear layer [29],

ReLU activation, and another linear layer. Residual connections ensure continuity across layers, enriching vectors with information from multi-head self-attention and position-wise feed-forward networks [30]. Four stacked encoder blocks constitute the encoder layers, with the outputs from the last block serving as input features for the subsequent pooling layer. A pooling layer is then introduced to reduce the spatial dimensions of the output volume. Finally, a feed-forward layer, commonly referred to as a multilayer perceptron, is utilized to regress and refine the output sequence.

Following the extraction of features through the introduced transformer architecture, subsequent tasks involve either classification. In the classification module, a softmax classification layer is applied to ascertain the insertion phase and tissue types, assigning corresponding labels.

## III. DATA COLLECTION, LABELING, LOW PASS FILTERING AND DATA AUGMENTATION

### A. Data Collection

For precisely measuring position tracking in a trajectory and the corresponding tissue reaction force, biopsy system with 6 degrees of freedom (DOF) consists of a potentiometric position sensor (IR robot: IR-10kΩ linearity potentiometer), a force measurement sensor (ATI model: Nano 43 Transducer), a biopsy cut needle (Bard instrument: 16-gauge/1.7mm). Robotic needle biopsy system can achieve 0.04mm precision under force/position control loop. These components work in tandem to replicate the conditions encountered during actual medical procedures, enabling a comprehensive assessment of the system's performance in a controlled laboratory environment.

The experimental setup for the robotic needle biopsy system involved the use of five distinct types of porcine tissues, including the liver, kidney, heart, belly, and hock tissues. The tissue sourcing from 5 adult pigs, irrespective of their sex, is provided by the UIUC Meat Science Laboratory. Tissue samples are then sectioned into uniform samples and placed in a controlled frozen environment ensuring the preservation of its natural structure and properties. The data collection begins with a careful defrosting of each tissue section, maintaining the tissue integrity. Each tissue type was essential for generating reliable and relevant data, covering a variety of real surgical scenarios from skin level to organ level needle penetrating. The entire biopsy procedure can be viewed as an integration of distinct, single tissue biopsies. Each individual tissue biopsy contributes a layer of information that, when combined, provides a comprehensive understanding of the tissue composition in the sampled area.

The robotic needle biopsy system then conducted a series of 50 distinct needle insertion procedures for each tissue section. These procedures were carried out with a constant feeding speed of 2mm/s, maintaining consistency in the needle insertion process. During each of these procedures, the system recorded each data frame for each procedure, including the corresponding reaction force exerted on the needle, the displacement of the needle within the tissue, and the data points of 20 Hz. Needle puncture timestamp is also recorded for labeling from surface penetration status.

### B. Data Labeling

The primary objective of the transformer classification model is to accurately detect transitions during needle-puncture events and classify the various tissue types involved. To train the transformer classification model, timestamps were carefully recorded to capture when critical events as the tissue punctures occurred. By applying the time-displacement profile, critical puncture events were located precisely with respect to the needle travel displacement.

An example of mixture labeling of events and tissue types after low-pass filtering is shown in Figure 4. During the initial phase of needle feeding, as indicated by both the blue curves in Figure 4, the tissue surface has not yet been penetrated. In this pre-puncture phase, it is inherently challenging to pinpoint the specific tissue type accurately. This is due to the lack of direct contact and interaction between the needle and the tissue surface. As the needle continues to feed into the tissue, a significant and identifiable event occurs—puncture. This is characterized by a notable drop in the interaction force, as illustrated by both the green curves in Figure 4. This phase is marked by the actual penetration of the needle into the tissue. Following the puncture phase, the needle interacts differently with the tissue. Friction and cutting forces become dominant, leading to distinct force dynamics. During this post-puncture phase, the data becomes more suitable for tissue type classification, as indicated by the red and purple curves in Figure 4. These labels correspond to specific tissue types encountered during the post-puncture phase, facilitating

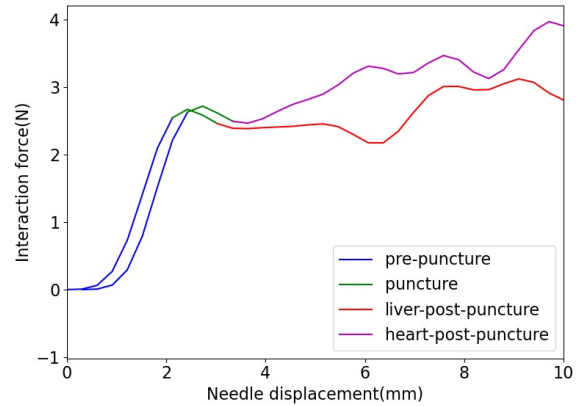

Fig. 4. Data labeling example of events and types for tissues

accurate classification based on the unique force profiles exhibited by each tissue type.

Thus, we create the following label types to meet the main aim to detect the transition in needle-puncture event and classify the different tissue types:

1. all tissues have a neutral phase.
2. all tissues have a pre-puncture phase.
3. all tissues have a puncture phase.
4. different labels (L for liver, K for kidney, H for heart, B for belly, and C for hock tissues) will be assigned for each post-puncture phase.

For example, a typical heart tissue would have a sequence of labels with respect to time as 1->1->1->2->2->2->3->3->H->H->H->1 during its needle insertion process.

*C. Low Pass Filtering and Data Augmentation*

Following the collection and labeling of data from the force and position channels, a 6th order Butterworth low pass filter [32] is used for enhancing data quality and usability for subsequent analysis. Here, the sampling rate for the force and position signals is set at 20Hz; hence, a cutoff frequency for the low pass filter is set to 10Hz. The cutoff frequency marks the point at which the filter begins to reduce the amplitude of high-frequency sensor noises while the low-frequency

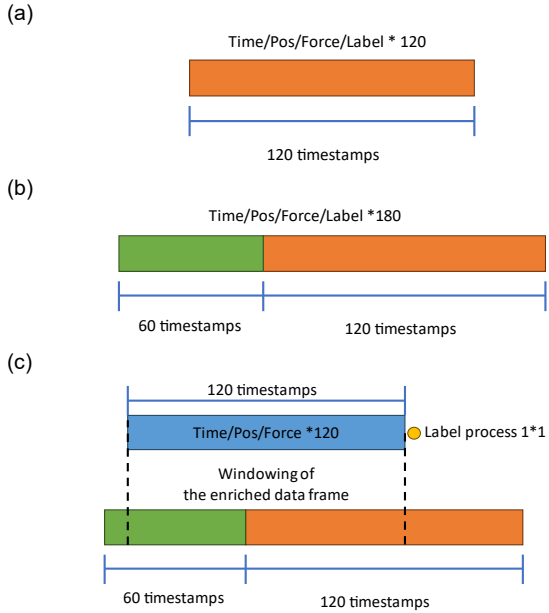

Fig. 5. Data augmentation. (a) Raw data frame. (b) Extended data frame. (c) Random windowing selection from extended data frame. interaction information is maintained.

For each of the 5 types of tissue, 8 sections are tested, with 50 repeated procedures on different locations on each section, resulting in 8(sections) * 50(procedures) * 5(types) = 2000 raw data frames. Each data frame comprises synchronized 120 timestamps, 120 position records, 120 force records, and 120 corresponding labels, as shown in Figure 5(a). For consecutive needle insertion involved with different tissues (i.e., belly then liver), force data could be stacked to illustrate the needle-tissue reaction during the procedure. We here separate the distinct reaction force with respect to each tissue type for a better clearance in training and classification.

To further enhance the dataset and match the potential cavities under skin during biopsy in abdominal or other areas, we applied a zero-padding enrichment technique, illustrated in Figure 5(b). Here, we added 60 timestamps and data points to the beginning of the raw data frame, with sequential timestamps, while setting both position and force to zero, to simulate the potential cavities or functional pause. The labels for these 60 timestamps were uniformly assigned as 1 (neutral/stop phase). Following this enrichment, we randomly selected 40 samples from the dataset, ensuring that their starting points were within the zero-padded region. However, the length of each sample remained at 120 timestamps. The label for each of these samples was designated as the last label in the new windowed data frame, as demonstrated in Figure 5(c).

As our aim is to build a state classifier for time series data, we consider the label corresponding to the last timestamp for a series of data points as the conclusive label. Therefore, we reduce the label from 120*1 into 1*1. By this data augmentation algorithm, we enrich the overall dataset from 2000 raw data frames into 80,000 data frames.

This enriched dataset provides adequate examples for the transformer classification method to learn from, allowing the model to capture various patterns and states within the time series data effectively. The increased dataset size contributes to the model's accuracy and generalizability, enhancing its ability to classify different states or conditions accurately. Moreover, the decision to employ a single conclusive label for the last timestamp greatly enhances the efficiency of the transformer model's training and classification processes. With a simplified label format, the model can process and learn from the data, leading to enhanced classification effectiveness.

## IV. MODEL TRAINING AND EVALUATION METRICS

For a comprehensive classifier, we need to have a transformer model trained on various tissues datasets. Our input data is represented as $X \in R^{B \times T \times F}$ and corresponding label $Y \in R^B$ in a batch, where $B$ stands for batch size, $T$ stands for timestamps in one data frame, and $F$ stands for features input to the model, which in our case F is 2 channels containing force and needle displacement. A linear projection and embedding layer are implemented to explicitly retain the order of signals in the time sequence. The output from the embedding layer is $E \in R^{B \times 2T}$. Next, we construct an encoder layer, and the output $E$ is fed into the multi-head attention module within one of the $N$ encoder blocks to extract valuable information. The multi-head attention module consists of $M$ heads, and each head processes the input independently.

For the $M$-th head, we construct query $Q_m$, key $K_m$ and value $V_m$ from input $E$, attention can be calculated as the following equation:

$$Attention = H_m(Q_m, K_m, V_m) = softmax\left(\frac{Q_m K_m^T}{\sqrt{d_k}}\right) V_m \quad (2)$$

where $K_m^T$ is the transpose of $K_m$, $\frac{1}{\sqrt{d_k}}$ is the scaling factor and $d_k$ is the dimension of $Q_m, K_m, V_m$. By concatenating the output of $M$ heads, we can obtain the aggregated temporal feature, $H$, for each data frame. To preserve the original features and prevent excessive smoothing during the classification process, we combine $E$ and $H$ then pass them through a layer-normalization layer, resulting in $L$. $L$ is then processed by a feed forward layer and then by a layer-normalization layer to obtain the output $O$ of one of the $N$ encoder blocks. We set number of heads as 8 and head size as 256. Four encoder blocks are contained in the encoder layer.

The transformer classification model is implemented with Keras on Tensorflow. The methodology implementation is in https://github.com/wangfanxin/tissue_classification. Using Kfold cross-validation technique, 80% of the dataset is set for training and testing, 20% of the dataset is set for evaluation. The evaluation of the tissue recognition algorithm relies on an accuracy metric. However, it is generally not possible to identify the tissue type before the puncture has taken place. In practical applications, our primary focus is on different stages of needle insertion process. Thus, we introduce 3 measures: accuracy for classifying pre-puncture phase as $A_{pre}$, accuracy for classifying puncture phase as $A_{punc}$, accuracy for classifying tissue types in post-puncture phase as $A_{tissue}$. The objective of the three metrics design is to assess the classifier's performance on the different stages of the needle insertion. Under $A_{tissue}$, distinct tissue classification precision is also recorded as $A_L, A_k, A_H, A_B, A_C$ for liver, kidney, heart, belly, and hock.

A series of comparative experiments were conducted to contrast the proposed approach with two well-established temporal classification modeling techniques [10], namely the RNN-LSTM model and the CNN model. This head-to-head comparison serves to elucidate the strengths and capabilities of the new method for tissue classification tasks.

All training, testing and offline validation is conducted on 8 * RTX 2080TI GPU and Intel(R) Xeon(R) CPU E5-2697A v4 @ 2.60GHz.

## V. RESULTS AND DISCUSSION

### A. Offline Classification Accuracy Evaluation

The average accuracy of the proposed classification algorithm is measured with 5 Kfold cross-validation, and the results for different models using metrics $A_{pre}, A_{punc}$, and $A_{tissue}$ are gathered in Table 1.

TABLE I
ACCURACY COMPARISON MEASURED ON VALIDATION SET.

| Method | Metric type | Acc [%] |
|---|---|---|
| Transformer | $A_{pre}$ | 95.10% |
|  | $A_{punc}$ | 94.58% |
|  | $A_{tissue}$ | 91.20% |
| RNN-LSTM | $A_{pre}$ | 92.27% |
|  | $A_{punc}$ | 90.11% |
|  | $A_{tissue}$ | 85.84% |
| CNN | $A_{pre}$ | 87.77% |
|  | $A_{punc}$ | 88.34% |
|  | $A_{tissue}$ | 81.56% |

Based on the data presented in Table 1, it is evident that the Transformer model consistently outperforms the other models across all metrics. During the pre-puncture phase, the Transformer model achieves an impressive accuracy of 95.3%, and for puncture phase and post-puncture phase, transformer achieved 94.2% and 91.7%, respectively. RNN-LSTM is the runner-up, it delivers reliable results with an accuracy of 92.27% during the pre-puncture phase. However, its performance in the puncture and post-puncture phases is not as impressive, achieving accuracies of 90.11% and 85.84%, respectively. CNN model performs the least effectively among the three models, yielding accuracy scores of 87.77% in the pre-puncture phase, 88.34% in the puncture phase, and 81.56% in the post-puncture phase.

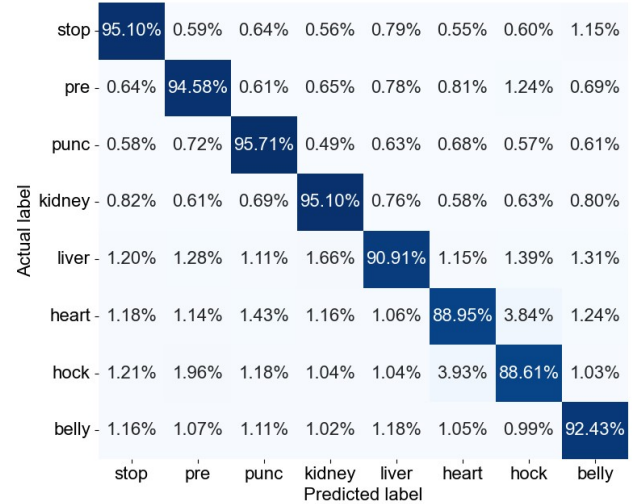

Fig. 6. Confusion matrix for accuracy obtained with the Transformer classification model.

The confusion matrix presented in Figure 6 offers a detailed view of the classification accuracy for the transformer model and errors in prediction.

A notable observation from the confusion matrix is the occasional misclassification of heart tissue as hock tissue. This misclassification can be primarily attributed to the similarities in muscle composition and mechanical properties between these two types of tissues. Both heart and hock tissues share comparable densities and textures, which can lead to challenges in differentiation during automated analysis.

On the other hand, for other tissue types, the classification accuracy of the system is impressively high, generally exceeding 92%. This high level of accuracy demonstrates the system's effectiveness in correctly identifying various tissues, such as liver, kidney, and belly tissues. Such a high success rate is indicative of the system's advanced capability in distinguishing between different tissue characteristics, which is essential for accurate diagnosis and treatment planning in medical procedures.

### B. Online Classification Accuracy Evaluation

We also developed an online classification system in which a needle was employed for biopsy procedures on liver tissue and heart tissue. The tissue types are intentionally kept unknown during the experiment. The needle is used to probe the tissue, and real-time data, including time, force, and needle displacement, are collected to form our input dataset. To facilitate real-time predictions, we integrate the most effective transformer model, based on our training results, into the experimental setup. To supply data to the model, a data frame consisting of 120 data points from the sensors is created and provided as input. In cases where the data frame contains fewer than 120 data points, we apply the same zero-padding technique introduced in Section III.C to ensure that the data frame reaches the desired length. We gathered data and ran all processing on CPU (Intel i9-13900H @ 2.600GHz) and GPU (NVIDIA GeForce GTX 4060) for the online tissue classification. In the case of the transformer model, the resulting inference time is below the targeted threshold of 10 ms, while the sampling rate of data collection is 20Hz (50ms).

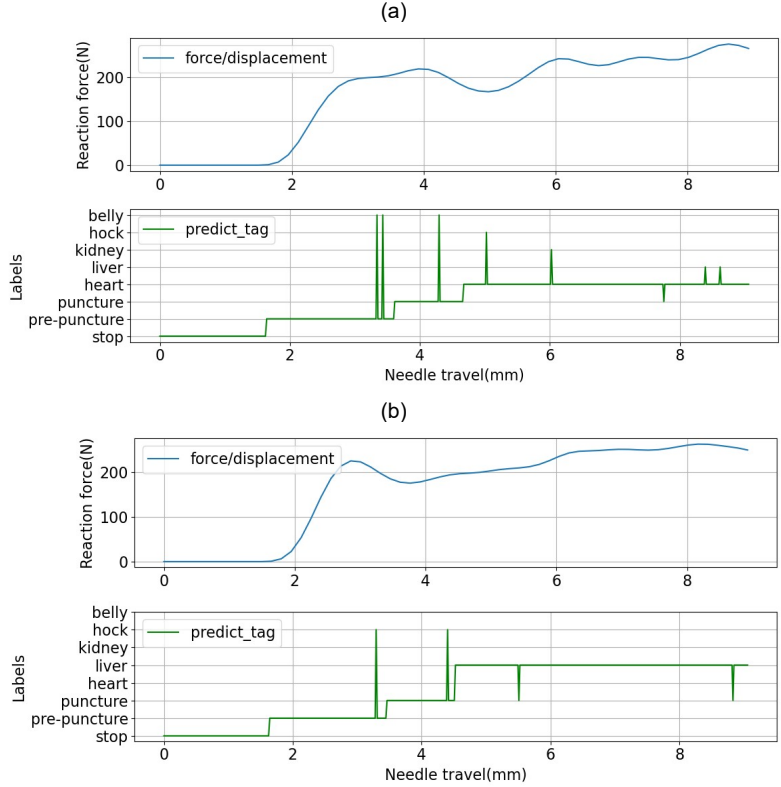

Fig. 7. Online classification result with needle force/displacement profile. (a) classification result with heart. (b) classification result with liver.

Model prediction result is illustrated in Figure 7, with the limitation of space, one heart case and one liver case are presented here. The force/displacement profile is depicted in blue, alongside the label generated from online classification is marked by green lines. The needle started from air for all tissue types, reaching the tissue surface with a constant speed of 2mm/s. The needle then fed in the region of interest, punctured into the surface with the same speed. From Figure 7, it is seen that for each tissue type, the pre-puncture and puncture phases were well detected. Very few error labels are generated during these periods. As the needle enters the post-puncture phase, distinct tissue labels are generated, and the overall classification accuracy is found to be satisfactory with more than 93%.

## VI. CONCLUSION AND FUTURE WORK

This paper introduces a novel approach that leverages the Transformer architecture to extract contextual information from time sequences of needle displacement and reaction forces, enhancing tissue classification. The key innovations and contributions are: (1) We present an advanced deep learning method based on the Transformer architecture, which proves highly effective in achieving tissue recognition. (2) Our approach involves detailed tissue modeling, categorizing the needle insertion process into three distinct phases, which ensures the accuracy and rationality of tissue classification, and allows for the systematic separation of tissues. (3) We perform real-time tissue classification on actual tissue samples, showcasing the practical applicability of our algorithm in real biomedical scenarios. In direct comparison with conventional temporal modeling-based frameworks, our proposed Transformer method consistently outperforms in tissue classification. The integration of our intelligence model and robotic biopsy system is depicted in Figure 8, the environment sensing and tissue transition detection offers a promising approach for robust and precise tissue classification, with broad applications in haptic biopsy sensing, ultimately enhancing the safety and effectiveness of medical procedures.

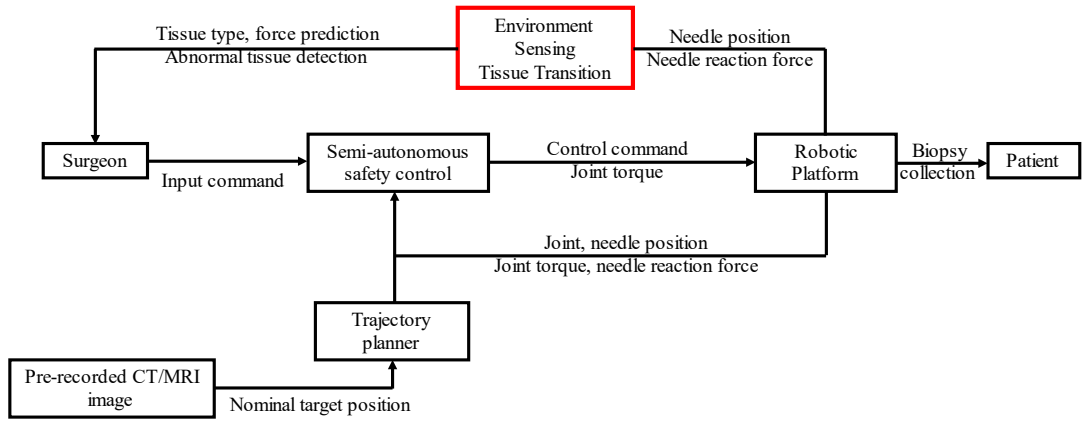

Fig. 8. Artificial intelligence implemented framework in robotic biopsy system.

Future work can be conducted on implementing tissue environment sensing in robotic biopsy or for automatic surgery, where reinforcement learning-based network could manipulate the robotic platform [23] and take charge of the needle insertion process with higher precision and accuracy.

Future work can also be conducted in the haptic guidance field. Firstly, by utilizing pre-recorded MRI, CT, or ultrasound images, a 3D reconstruction of a specific region of interest and its surroundings can be created, resembling electroanatomic reconstruction [24-25]. On the other hand, haptic guidance can be employed to detect changes in tissue compositions [26-27]. A more detailed classification model can be developed and trained with needle interaction measured on tissues separated by compositions, or combined tissue but layer transitions are carefully recorded. With further training on healthy and diseased tissue, the work developed here can be extended in new directions to significantly enhance the efficacy and precision of biopsy procedures.